# Modelling Optimal Policies of Demand Responsive Transport and Interrelationships between Occupancy Rate and Costs

Jani-Pekka Jokinen
Max Planck Institute for Dynamics and Self-Organization
Am Faßberg 17, 37077 Göttingen, Germany. (jani-pekka.jokinen@ds.mpg.de)

**Abstract**

This paper presents a model addressing welfare optimal policies of demand responsive transportation service, where passengers cause external travel time costs for other passengers due to the route changes. Optimal pricing and trip production policies are modelled both on the aggregate level and on the network level. The aggregate model is an extension from Jokinen (2016) with flat pricing model, but occupancy rate is now modelled as an endogenous variable depending on demand and capacity levels. The network model enables to describe differences between routes from the viewpoint of occupancy rate and efficient trip combining. Moreover, the model defines the optimal differentiated pricing for routes.

Keywords: demand responsive transport, occupancy rate, optimal pricing, social welfare

# 1 Introduction

Automated demand responsive transport (DRT) and other new flexible micro transport services (FMTS) are emerging due to the technological advancements in information technology and intelligent transport technology (Jokinen et al., 2017), (Weckström et al., 2018). These services can provide shared trips without transfers from stop-to-stop or even from door-to-door. Depending on decided transport policies, the DRT services can be either competitors for regular public transport or alternatively complementary services making public transport service provision more comprehensive and attractive compared to private car.

An economic analysis of demand responsive transportation has been conducted in several studies during last decade. Jokinen et al. (2011) analyzed cost-effectiveness of DRT with simulations. Wang et al. (2014) analyzed demand and future possibilities of DRT in the UK. Ryley et al. (2014) studied the contribution of DRT to a sustainable local public transport system. Davison et al. (2014) studied the provision of DRT in the UK and conducted a national survey of DRT providers. Jokinen (2016) presented analytical model of automated DRT and shared taxi services and defined welfare optimal pricing and trip production policies. One conclusion from the previous studies is that occupancy rate of DRT vehicles is an essential indicator describing efficiency of trip combining and more detailed analysis is needed for understanding interrelationships between occupancy rate, costs and optimal DRT policies.

In this paper, we analyze welfare optimal DRT policies and interrelationships between occupancy rate and costs (both internal and external costs) by extending analytical DRT model presented by Jokinen (2016), where occupancy rate was given as exogenous variable. We present a model where occupancy rate is defined endogenously by other variables describing the state of the DRT system, i.e., trip demand and available fleet capacity. Moreover, we analyze the model on the network level for taking into account the network aspects as different road networks enable variable possibilities for detours influencing on the expected and realized travel time externalities that new passengers cause for other passengers in DRT vehicles. Furthermore, we present numerical examples of the model for illustrating its use in policy analysis and pricing.

In the following section, we present a model for trip demand and DRT service, and we analyze interrelationships between occupancy rate and costs. Moreover, we model welfare optimal policies for trip production and pricing of DRT. In Section 3, we consider optimal policies for variable trip distances on the road network. In Section 4, we present numerical examples of the model. Finally, in Section 5, some conclusions are drawn and we discuss policy implications of the results.

## 2 Model

The demand for DRT trips within a limited service area is assumed to be a decreasing function of the generalized costs, defined by a demand function:

$$X = D(G), \frac{dD}{dG} < 0 \qquad (1)$$

where X is the number of passengers traveling by DRT during the considered time period, e.g., one hour or day. G is the average generalized cost, including passenger money and time costs defined by equation:

$$G = p + q + k \qquad (2)$$

where $p$ is the ticket price, $q$ is the average value of waiting time, and $k$ is the average value of in-vehicle time.

The waiting time for a DRT trip is defined by the time the dispatched vehicle needs to drive to the pick-up point, similarly to call taxes, but with more complex routing due to the simultaneously served customers. An increase in the number of operating DRT vehicles in the service area decrease the expected distance between the dispatched vehicle and the pick-up point, and thereby decrease the average waiting time. Respectively, new additional passengers increase the expected route length to the pick-up point and the expected number of stops before the pick-up point, which increase average waiting time. Therefore, we define the average value of waiting time, $q$, as a function of the trip production, R, and the demand, X:

$$q = Q(R,X), \frac{dQ}{dR} < 0, \frac{dQ}{dX} > 0 \quad (3)$$

Variable R measures trip production as a number of potential seat kilometers supplied, i.e., R is a product of potential vehicle kilometers and number of seats per vehicle. The distinction between potential and actual seat kilometers is essential, because a DRT vehicle is not necessarily moving if it's not temporarily dispatched to any trip request similarly as taxis waiting new customers.

In the DRT service, other passengers increase the value of travel time cost by increasing the route length, number of stops and crowding in the vehicle. These travel time costs can be reduced by increasing the trip production capacity relative to the demand, which consequently decreases the average occupancy rate. Moreover, the average value of travel time depends on the average trip distance, $a$. Therefore, the average value of travel time is modelled as a function of trip production capacity, demand, and average trip distance:

$$k = K(R,X,a), \frac{dK}{dR} < 0, \frac{dK}{dX} > 0, \frac{dK}{da} > 0 \quad (4)$$

where the signs of the partial derivatives are based on the reasoning above.

We define the inverse demand function by inverting the function (1) and using equation (2):

$$p + q + k = G = D^{-1}(X), \frac{dD^{-1}(X)}{dX} < 0 \quad (5)$$

The collective willingness to pay of passengers during the considered time period can be written by using equations (2) - (5):

$$W = \int_0^X D^{-1}(x)dx - [Q(R,X) + K(R,X,a)]X \quad (6)$$

where the first term is the collective gross willingness to pay as a function of the demand level, and the second term is the collective time cost of passengers.

Total costs of DRT operations, C, are given by the linear function:

$$C = c_0 + c_1 F + c_2 R + c_3 X + c_4(Xa + Xb) + c_5(Xc) \tag{7}$$

where $c_0$ is the fixed cost of the DRT operator, that is, costs which are not dependent on the realized demand and provided fleet capacity. The parameter $c_1$ is a cost of capital required for capability to produce 1 seat kilometer during considered time period which is multiplied by the variable F denoting the quantity of the capital. The parameter $c_2$ is a cost of operations required for capability to produce 1 seat kilometer during considered time period, which is multiplied by the variable R denoting the capacity of the operating fleet. The parameter $c_3$ is a distance-independent cost for an additional customer, that is, costs due to dispatching and additional stops for boarding and alighting from the vehicle. The average trip distance, a, is based on the shortest route, and the term $Xa$ is the number of served passenger kilometers excluding detour kilometers, b, from the term. Thus, the term $(Xa + Xb)$ is the total kilometers travelled by passengers in vehicles, and the parameter $c_4$ is a route-independent cost of passenger traveling 1 kilometer in a DRT vehicle, including costs for increase in fuel consumption due to the extra weight, contamination, and abrasion of seats. The last term $c_5(Xc)$ is the cost of driven kilometers by DRT vehicles. The parameter $c$ is an average increase in a route length of vehicle due to an additional trip. The term $Xc$ gives the total kilometers driven by the DRT fleet. The parameter $c_5$ is a cost of kilometer driven by an empty vehicle. As in Jokinen (2016) a cost of driving an empty vehicle is used in this term, because the weight of customers (and luggage) are taken into account in the previous term, $c_4(Xa + Xb)$.

A feasible level of operations required for potential seat kilometers, R, is constrained by the quantity of operator's capital, that is, the fleet size and related equipment, measured by the variable F:

$$R \leq UF \tag{8}$$

where U is an upper limit for the utilization rate of capital, which is normally below 1 (and 0,8 in the numerical examples of the section 4).

A feasible level of passenger kilometers, $Xa$, during the considered time period is constrained by the level of operations required for potential seat kilometers, R:

$$aX \leq O(X,R)R \tag{9}$$

where O(X, R) is an upper limit for the occupancy rate of vehicles as a function of X and R, which is normally below 1, and it is often below 0,1 (Jokinen, 2016, Jokinen et al., 2017).

An occupancy rate (for a certain time period) is defined as the produced trip kilometers (excluding detours) divided by the potential seat kilometers of the fleet:

$$O = \frac{Xa}{R},$$

which can have values between 0 and 1, where the value 1 represents an ideal use of the fleet meaning that there are no idle minutes, nor detours or empty seats.[1] Thus, we have a clear intepretation for the value 1 of the occupation rate, but it is very difficult (or impossible) to reach even with private taxi fleets not to mention DRT services where detours can be seen as enablers of trip combining. Therefore the maximum value (1) of the occupancy rate is not feasible for DRT in practice (and it is not even desireble as detuours are usually necessary for trip combining). However, an increase in occupancy rate means more effifient use of the fleet, and therefore, for modelling and analysing optimal pricing and trip procuction of DRT fleets it is important to understand and identify the various reasons that cause reductions in occupancy rate.

---

[1] Other intuitively reasonable measure for occupancy rate could be $\frac{X(a+b)}{R}$ as it describes the ratio of occupied seat kilometers and potential seat kilometers, but this measure has not as clear intepretation for value 1, and increase of this measure does not mean unambiguosly more efficient use of the fleet as it can be achieved by increasing only detours, b.

We have identified 5 technical reasons which reduce occupancy rate of DRT fleet. Firstly, there are empty seats (at least temporarily) in the vehicles that produce trips for number of passengers less than the seat capacity. The effect of this reason could be reduced by increasing passenger waiting times (or time windows), like in many DRT/shared taxi services in the African countries where the trip will not begin until the vehicle is full. The second reason are detours, i.e., even the vehicle is full the detour, b, causes that produced trip kilometers (based on shortest routes) are less than potential seat kilometers, thus Xa < R, and consequently O < 1. The third reason are driven kilometers only for pickup, which causes vehicles to drive empty or increasing detours for other passenger already in the vehicle if the pickup route and the shortest route of the first customer do not match perfectly. The fourth reason is the idle time of vehicles, which can be relatively high outside the peak hours. The fifth reason is the time used for picking up and dropping off passengers, which can be relatively high for special passenger groups requiring assistance.

The social planner's objective is to maximize the welfare defined as the passengers' willingness to pay, defined in equation (6), minus the transport operator's costs, defined in equation (7), with regard to X, R, and F, subject to the restrictions in (8) and (9). The Lagrangean function for the maximization problem can be formulated as:

$$L = W - C - \lambda(aX - O(X,R)R) - \mu(R - UF) \tag{10}$$

The following first-order conditions for optimal X, R, and F can be derived:

$$\frac{\partial L}{\partial X} = G - q - k - \frac{\partial Q}{\partial X}X - \frac{\partial K}{\partial X}X - (c_3 + c_4 a + c_4 b + c_5 c) - \lambda(a - R\frac{\partial O}{\partial X}) = 0 \tag{11}$$

$$\frac{\partial L}{\partial R} = -\frac{\partial Q}{\partial R}X - \frac{\partial K}{\partial R}X - c_2 + \lambda(O + \frac{\partial O}{\partial R}R) - \mu = 0 \tag{12}$$

$$\frac{\partial L}{\partial F} = -c_1 + \mu U = 0 \tag{13}$$

If capital and operations are fully utilized the constraints (8) and (9) are equalities, and optimal p, R, and F can be deduced from equations (11) - (13):

$$p = \frac{\partial Q}{\partial X}X + \frac{\partial K}{\partial X}X + c_3 + c_4 a + c_4 b + c_5 c +$$

$$(c_1/((O + \frac{\partial O}{\partial R}R)U) + c_2/(O + \frac{\partial O}{\partial R}R) + \frac{\partial Q}{\partial R}X/(O + \frac{\partial O}{\partial R}R) + \frac{\partial K}{\partial R}X/(O + \frac{\partial O}{\partial R}R))(a - R\frac{\partial O}{\partial X}) \tag{14}$$

$$aX = O(X,R)R = O(X,R)UF \tag{15}$$

The term $\frac{\partial Q}{\partial X}X$ is the increase in the waiting time costs the passenger causes for other passengers. The term $\frac{\partial K}{\partial X}X$ is the increase in the in-vehicle travel time costs the passenger causes for other passengers, which is usually higher in the DRT service than in regular public transport, because a new passenger in DRT usually causes route changes, which increases travel time cost of other passengers more than just due to the crowding. The terms $\frac{\partial Q}{\partial R}X$ and $\frac{\partial K}{\partial R}X$ describe savings in travel time costs and waiting time costs due to a marginal increase in route production. In equation (14), the optimal price is equal to the marginal costs of operator (the terms 3-6 and the first two terms in the parenthesis in equation (14)) plus external costs to the passengers (the terms 1-2) minus value of travel time savings from increased route production (the last two terms in parenthesis, i.e., $\frac{\partial Q}{\partial R}X/(O + \frac{\partial O}{\partial R}R)$ and $\frac{\partial K}{\partial R}X/(O + \frac{\partial O}{\partial R}R)$). The optimal occupancy rate, $O^*$, is defined as a function of X and R, i.e., $O^* = O(X,R)$. Several studies have shown that trip combining possibilities and occupancy rate are increasing with the scale of the service (Jokinen, 2011).

The presented model treats travel distances only as averages. Thus, the optimal price (14) can be interpreted as the optimal flat rate. Flat rates are often applied in regular public transport with fixed routes and schedules, but sometimes also in DRT services, e.g., The Ecobus service in Germany. However, typically in DRT services, pricing is based on the trip distance, because marginal costs of DRT trips are strictly dependent on distance, almost similarly to taxi trips. In addition to a trip distance, a structure of the road network and demand densities on different parts of the network have effects on the social marginal costs of the produced trips. Next, we consider these effects on optimal pricing by modifying the model for taking into account road network and allowing price differentiation based on a trip origin and destination.

## 3 Optimal Policies on the Road Network

As previously, the demand for DRT trips is assumed to be a decreasing function of the generalized costs:

$$X = \sum_r X_r = \sum_r D_r(G_r), \frac{dD_r}{dG_r} < 0, \quad r = 1, 2, \ldots, n. \tag{16}$$

Where $n$ is a number of routes in the road network, X is the total demand for the DRT trips, which is a sum of demands for each routes, $\sum_r X_r$. $G_r$ is an average generalized cost of a route, r, including ticket price, waiting time and in-vehicle time costs as previously, i.e., $G_r = p_r + q_r + k_r$.

Waiting time on the routes of the service network is defined as a function R and demand on the routes:

$$q_r = Q_r(R, X_1, X_2 \ldots X_n), \frac{dQ_r}{dR} < 0, \frac{dQ_r}{dX_r} > 0, \frac{dQ_r}{dX_i} > 0, \frac{dQ_r}{dX_i^c} < 0. \tag{17}$$

Where partial derivative of waiting time with respect to R is negative as in the aggregate model. The partial derivative of waiting time with respect to demand is positive similarly to the aggregate model, except in the case of demand for complement routes, $X_i^c$, which have destinations near the origin on the route r. Therefore, increase of demand for the complement route increase number of vehicles available for the route r, and consequently $\frac{dQ_r}{dX_i^c} < 0$.

As previously in the aggregate model, the average value of travel time on a route is a function of trip production, demand, and trip distance of the route (instead of average distance):

$$k_r = K_r(R, X_1, X_2..X_n, d_r), \frac{dK_r}{dR} < 0, \frac{dK_r}{dX_r} > 0, \frac{dK_r}{dd_r} > 0, \quad (18)$$

where the signs of the partial derivatives are based similar to the aggregate model. Moreover, the value of the in-vehicle time is naturally increasing with the route distance, $d_r$.

By using equations 16 - 18, the inverse demand function and collective willingness to pay can be defined for route r:

$$p_r + q_r + k_r = G_r = D^{-1}(X_r), \frac{dD^{-1}(X_r)}{dX_r} < 0. \quad (19)$$

$$W_r = \int_0^{X_r} D^{-1}(X_r)dx - [Q_r(R, X_1, X_2..X_n) + K_r(R, X_1, X_2..X_n, d_r)]X_r, \quad (20)$$

where the first term is the collective gross willingness to pay as a function of the demand level, and the second term is the collective time cost of passengers.

Total costs, C, are given by the linear function:

$$C = c_0 + c_1 F + c_2 R + c_3 X + \Sigma_r(c_4(X_r d_r + X_r b_r) + c_5 X_r c_r), \quad (21)$$

where the three first terms are similar to the aggregate model, and the three last terms are presented for each route and then summed over the routes.

A feasible level of seat kilometers, $\Sigma_r X_r c_r * s$, during the considered time period is constrained by the level of operations required for potential seat kilometers, R:

$$\Sigma_r X_r c_r * s \leq R \quad (22)$$

where s is the number of seats in the vehicle and $c_r$ is the average additional vehicle kilometer per new passenger on the route r.

As in the aggregate model, the social planner's objective is to maximize the welfare defined as the passengers' willingness to pay, defined in equation (20), minus the transport operator's costs, defined in equation (21), with regard to X, R, and F, subject to the restrictions for R and F. The Lagrangean function for the maximization problem can be formulated as:

$$L = \Sigma_r(W_r) - C - \lambda(\Sigma_r(X_r c_r * s) - R) - \mu(R - UF)$$

$$= \Sigma_r \left\{ \int_0^{X_r} D^{-1}(X_r) dx_r - [Q_r(R, X_1, X_2..X_n) + K_r(R, X_1, X_2..X_n, d_r)] X_r \right\} - C$$

$$-\lambda(\Sigma_r(X_r c_r * s) - R) - \mu(R - UF) \quad (23)$$

The first order conditions:

$$\frac{\partial L}{\partial X_r} = G_r - q_r - k_r - \Sigma_i \frac{\partial Q_i}{\partial X_r} X_i - \Sigma_i \frac{\partial K_i}{\partial X_r} X_i - (c_3 + c_4 d_r + c_4 b_r + c_5 c_r) - \lambda(c_r * s) = 0 \quad (24)$$

$$\frac{\partial L}{\partial R} = -\Sigma_i \frac{\partial Q_i}{\partial R} X_i - \Sigma_i \frac{\partial K_i}{\partial R} X_i - c_2 + \lambda - \mu = 0 \quad (25)$$

$$\frac{\partial L}{\partial F} = -c_1 + \mu U = 0 \quad (26)$$

If capital and operations are fully utilized the constraints for R and F are equalities, and optimal p, R, and F can be deduced from equations (24) - (26):

$$p_r = \Sigma_i \frac{\partial Q_i}{\partial X_r} X_i + \Sigma_i \frac{\partial K_i}{\partial X_r} X_i + (c_3 + c_4 d_r + c_4 b_r + c_5 c_r) + (c_1/U + c_2 + \Sigma_i \frac{\partial Q_i}{\partial R} X_i + \Sigma_i \frac{\partial K_i}{\partial R} X_i)(c_r * s) \quad (27)$$

$$\Sigma_r(X_r c_r * s) = R = UF \quad (28)$$

The terms in equation (27) have the same intepretation as in equation (14) except that production costs and external costs are now for route r and the capacity constraint is defined differently.

## 4 Numerical Examples

In this section, we present numerical examples for illustrating the model and it's use in transport policies of DRT services. We specify demand, X, as the log-linear function of generalized cost, G:

$$X = d_0 G^{-d_1} \quad (29)$$

Following Jokinen (2016), we assume the value of the demand elasticity with respect to generalized cost as 1.5, that is, $d_1$=1.5. With the assumed demand elasticity, hourly trip demand and generalized costs, the parameter $d_0$ of the demand function can be estimated. The values of the parameters and variables in the numerical examples are presented in Table 1.

**Table 1**

**Values of Parameters and Variables in the Numerical Examples[2]**

| Parameters / Variables | Aggregate model | Route AB | Route BC | Route AC | Route ABC |
|---|---|---|---|---|---|
| Demand (base value), per hour | 100 | 33,33 | 33,33 | 16,67 | 16,67 |
| Elasticity, $d_1$ | 1.5 | 1.5 | 1.5 | 1.5 | 1.5 |
| $d_0$ | 3340 | 1113 | 778 | 638 | 690 |
| Value of time (€ per minute) | 0.2 | | | | |
| Average waiting time | 6 | | | | |
| Average value of waiting time, q | 1.20 | | | | |
| Average in-vehicle time | 19 | | | | |
| Average value of in-vehicle time, k | 3.80 | | | | |
| Price, p | 5.37 | | | | |
| Trip distances (km) | 6.38 | 7.14 | 3 | 9 | 10.14 |
| Detour (km) | 0,19 | 0 | 0 | 0 | 1.14 |
| Average additional vehicle km per trip, c or $c_r$ | 4.104 | 3 | 2.312 | 8 | 6 |
| Marginal waiting time externality, $\frac{\partial Q}{\partial X}$ | 0.003 | | | | |
| Marginal in-vehicle time externality, $\frac{\partial K}{\partial X}$ | 0.0009 | | | | |
| Marginal waiting time external effect of fleet capacity, $\frac{\partial Q}{\partial R}$ | -0.00008 | | | | |
| Marginal in-vehicle time external effect of fleet capacity, $\frac{\partial K}{\partial R}$ | -0.000007 | | | | |
| Utilization rate of capital, U | 0.8 | | | | |
| Occupancy rate (simulated base value), O | 0.155 | | | | |
| Marginal occupancy rate external effect of demand, $\frac{\partial O}{\partial X}$ | 0.00003 | | | | |
| Production costs | | | | | |
| Cost of capital per seat kilometer, $c_1$ | 0.006 | | | | |
| Cost of operations per seat kilometer, $c_2$ | 0.098 | | | | |
| Distance-independent cost of passenger, $c_3$ | 0.016 | | | | |
| Route-independent cost of passenger per kilometer, $c_4$ | 0.005 | | | | |
| Cost of kilometer driven by an empty vehicle, $c_5$ | 0,5 | | | | |

The road network in the numerical example is based on the four stops (A, B, C and D) and four trip routes available for passengers, consisting of the three direct trips from A to B (AB), B to C (BC), A to C (AC) and one trip with detour from A to C through B (ABC). The specified demand functions for the trips and related distances and parameters are also presented in Table 1 and the road network is presented in Figure 1.

---

[2] The values of the aggregate model are adopted from (Jokinen, 2016), except demand level and occupancy rate that are based on the simulations presented in (Jokinen et al., 2011).

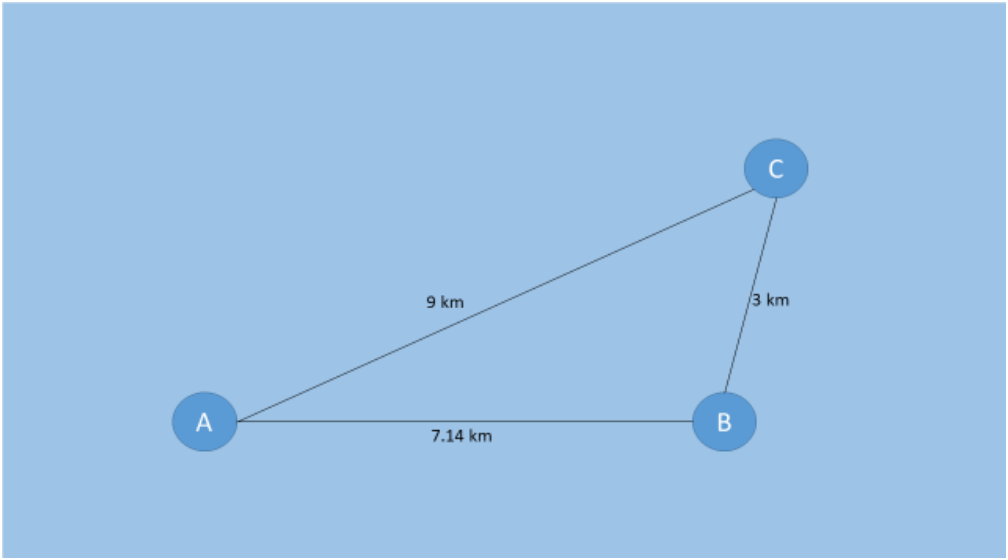

**Figure 1. The road network of the numerical example.**

Table 2 presents the values of the aggregate model for the base case and for optimized price (p) and capacity (R ), where the occupancy rate is defined as a function of demand and capacity (equation 9), which partial derivative with respect to demand is positive based on the simulation results of Jokinen et al., 2011.

**Table 2**

**Numerical Example for the Aggregate Model**

| Variables | Base values | Optimal (p, R) |
|---|---:|---:|
| Demand, X (base value) | 100 | 102 |
| Average value of waiting time, Q | 1.20 | 1.99 |
| Average value of in-vehicle time, K | 3.80 | 3.80 |
| Price (flat / weighted average), P | 5.37 | 5.24 |
| Occupancy rate | 0.1550 | 0.1555 |
| Quantity of operations, R | 4104 | 4180 |
| Operation costs, C | 507 | 517 |
| Collective willingness to pay | 2609 | 2620 |
| Social welfare | 2102 | 2103 |

As can be seen from Table 2, the optimized price is lower and provided capacity is somewhat higher leading to higher operational cost, but the demand and passenger benefits (measured by the collective willingness to pay) are respectively higher.

Table 3 presents respectively the values of the network model for the base case and for optimized trip prices ($p_r$) and capacity (R ), where prices for each route are differentiated based on the additional vehicle kilometer that each trip request on average causes (parameter $c_r$ in Table 1). In practice, the parameter $c_r$ can be calculated in modern DRT services from the vehicle dispatching system. In this example, we assumed the values of $c_r$ for describing a situation where majority of trips are requested for trips AB and BC, and therefore, the trip ABC with detour causes on avarage less additional vehicle kilometers ($c_{ABC} = 6$) than the direct route AC ($c_{AC} = 8$), because the trip ABC can be often combined with the trips AB and BC.

**Table 3**

**Numerical Example for the Network Model**

| Variables | Base values | Optimal ($p_r$, R) |
|---|---|---|
| Demand, X | 100 | 101.5 |
| $X_{AB}$, 7,14 km | 33.333 | 36.2 |
| $X_{BC}$, 3.00 km | 33.333 | 45.8 |
| $X_{AC}$, 9.00 km | 16.667 | 8.4 |
| $X_{ABC}$, 10,14 km | 16.667 | 11.1 |
| Average value of waiting time, Q | 1.20 | 1.25 |
| Average value of in-vehicle time, K | 3.50 | 3.11 |
| $K_{AB}$, 7.14 km | 3.80 | 3.81 |
| $K_{BC}$, 3.00 km | 1.60 | 1.60 |
| $K_{AC}$, 9.00 km | 4.79 | 4.80 |
| $K_{ABC}$, 10,14 km | 5.40 | 5.41 |
| Price, p | 5.37 | 5.44 |
| $p_{AB}$ | 5.37 | 4.76 |
| $p_{BC}$ | 5.37 | 3.75 |
| $p_{AC}$ | 5.37 | 11.94 |
| $p_{ABC}$ | 5.37 | 9.08 |
| Quantity of operations, R | 4104 | 3478 |
| Total costs, C | 631 | 535 |
| Collective willingness to pay | 2550 | 2509 |
| Social welfare | 1920 | 1974 |

As can be seen from Table 3, the optimal prices are higher for longer distances and especially higher for trips causing more additional vehicle kilometers, i.e., the price of trip AC is higher than for the trip ABC, even though the direct distance between the origin (A) and destination (C) is the same (9 km) and the travelled kilometers in the trip ABC are higher due to the detour. Moreover, the optimized capacity, R, is now lower than in the base case.

# 5 Conclusion and Discussion

We have presented a model for demand responsive transportation, which defines welfare optimal prices and trip production. The first version of the model describes a DRT market on the aggregate level with endogenous occupancy rate. The second model describes a DRT market on the network level and presents alternative way to present capacity constraint for trip production by using average additional vehicle kilometers for routes. These vehicle kilometers can be measured in modern DRT services (i.e., automated DRT services) from dispatching systems. The measured values enable to analyze relative efficiency of the routes on the road network and to define optimal differentiated prices for the routes.

As we already stated, new automated DRT services can become either competitors or complementary services for regular public transport. The former case can lead to increased motor vehicle kilometers and consequently to increased negative externalities of transport (such as air pollution, noise and congestion), especially if the occupancy rate of the DRT fleet is relatively low. In the latter case the effects on the negative externalities can be opposite, but only if it attracts passengers from transport modes with lower average occupancy rates (e.g., private cars and taxis) instead of more sustainable transport modes. Welfare optimal pricing and trip production policies play crucial role in achieving the objectives for efficient use of DRT services as a part of sustainable transport systems. The two main types of DRT services are (1.) door-to-door services with individual passenger stops and (2.) stop-to-stop service based on predefined available stops that are typically the same than for regular public transport. The first type is attractive especially for passenger groups avoiding walking long distances, e.g., for health or safety reasons. The second type, on the other hand, enables more efficient trip production due to the better trip combing possibilities. DRT operators could basically provide the both service types simultaneously and thereby satisfy larger passenger groups with varying travel needs and preferences. The average additional vehicle kilometers are higher for door-to-door trips, and therefore, optimal prices would be higher than for stop-to-stop trips as the presented network model and the related numerical example show. The same applies also if DRT is provided as a feeder service for regular public transport, i.e., price of DRT trip from a bus stop to the destination with high demand density (e.g., railway station) should be relatively lower than the price from the same bus stop to the passenger's home address in the area with low demand density.

# Acknowledgements

We acknowledge the funding of the Horizon 2020 program for the Ecointels project. We thank the Max Planck Institute for Dynamics and Self-Organization for proving research facilities. We also thank The Next Generation Mobility research group for the help and useful discussions that have influenced this paper.